# A Synergistic Approach to Wildfire Prevention and Management Using AI, ML, and 5G Technology in the United States

Okoro, C. Stanley[1], Lopez Alexander[2], Unuriode, O. Austine[3]

[1,2,3,4] Department of Computer Science, Austin Peay State University, Clarksville, USA.
STAN.OKORO@GMAIL.COM, TOLUWANILOPEZ@GMAIL.COM, AUSTINEUNURIODE@GMAIL.COM

## ABSTRACT

Over the past few years, wildfires have become a worldwide environmental emergency, resulting in substantial harm to natural habitats and playing a part in the acceleration of climate change. Wildfire management methods involve prevention, response, and recovery efforts. Despite improvements in detection techniques, the rising occurrence of wildfires demands creative solutions for prompt identification and effective control. This research investigates proactive methods for detecting and handling wildfires in the United States, utilizing Artificial Intelligence (AI), Machine Learning (ML), and 5G technology. The specific objective of this research covers proactive detection and prevention of wildfires using advanced technology; Active monitoring and mapping with remote sensing and signaling leveraging on 5G technology; and Advanced response mechanisms to wildfire using drones and IOT devices. This study was based on secondary data collected from government databases and analyzed using descriptive statistics. In addition, past publications were reviewed through content analysis, and narrative synthesis was used to present the observations from various studies. The results showed that developing new technology presents an opportunity to detect and manage wildfires proactively. Utilizing advanced technology could save lives and prevent significant economic losses caused by wildfires. Various methods, such as AI-enabled remote sensing and 5G-based active monitoring, can enhance proactive wildfire detection and management. In addition, super intelligent drones and IOT devices can be used for safer responses to wildfires. This forms the core of the recommendation to the fire Management Agencies and the government.

1. **INTRODUCTION**

In the past decades, wildfires have become one of the main global environmental issues causing havoc in areas such as tropical Amazon and cold Serbia [49]. Recent wildfires in Australia, Indonesia, Greece, the United States, and Brazil have not only damaged ecosystems but have also caused climate change through carbon emissions [32, 55]. Globally, it is estimated that 3.53 Pg. of carbon is emitted by fires annually, which is 25-35% of the total net carbon emission [27]. A rise in world temperature by 2°C has led to increased wildfire frequency. Only 3% of wildfires have occurred naturally with most of them being triggered by anthropogenic activities or human-induced factors [13, 55] Natural causes include spontaneous combustion, volcanic activities, and lightning strikes. The anthropogenic activities include negligence, uncontrolled agricultural activities, and land changes [7].

Over the past four decennials, burned regions from wildfires in the United States have roughly quadrupled [34]. This swift growth has been propelled by various factors including fuel accumulation owing to an effect of fire suppression over the last centennial and a more recent upsurge in fuel aridity in the western US. This trend is expected to continue due to increasing warm climates [3]. In the contiguous United States (CONUS) wildfires between 2011 and 2016 led to crop and property damage valued at $ 3.5 billion, suppression efforts valued at $ 12.4 billion, and the loss of 162 lives [19]. The United States experiences wildfires during the four weather seasons. During winter, wildfires are mainly found in the Southeast. As spring approaches, wildfire detectors move northwards due to increased fires across the central US. During summer, the wildfire peaks in the western US. During fall, California, and the Southeast US experience wildfires [20].

To effectively address the wildfire threat, comprehensive fire management strategies are fundamental. Early detection and rapid response systems play an important role in monitoring and surveillance, facilitating timely action [7]. Fire prevention measures such as fuel management, controlled burns, and public awareness and educational campaigns are important in minimizing fire risks. Fuel management and controlled burning practices alleviate fire risk while public awareness and community engagement programs encourage safe practices and early reporting of fire threats. Techniques of fire suppression entail the deployment of firefighting infrastructure and resources to contain and put off fires. Active capacity building and community involvement improve fire-safe practices and community resilience [3].

Early detection of wildfires is vital to their management and monitoring [49]. Surveillance through watchtowers, aircraft, drones, and traditional physical sensors has proven to be inadequate in fire detection [42, 61]. Aerial, terrestrial, and satellite devices have been mainly utilized to detect wildfires at the initial stages [37]. With satellites, it is possible to detect extensive wildfires using thermal infrared and high-temperature-sensitive short-wave infrared channels [25]. Satellite devices, owing to their affordability and extensive coverage, are valuable for near-real-time detection, monitoring, and evaluation of wildfire-affected areas. [49]. However, despite exhibiting a high spatial resolution, sun-synchronous satellites have a low time resolution therefore making it challenging to detect wildfires in real time. Similarly, it may be challenging to detect wildfires at the initial stages using geostationary satellites as they display low spatial resolution and high time resolution.

With the rapid progress in image processing technologies and digital cameras, deep learning object detection algorithms can utilize graphic cards and parallel computing to achieve near-real-time processing speeds. There is a growing global interest in creating real-time models for wildfire detection through the application of common video-based surveillance, employing deep learning technologies like convolutional neural networks (CNNs). [42, 51]. CNNs are computer vision techniques that display many benefits over traditional smoke and flame detection due to their flexible system installation, high accuracy, early fire detection, and ability to detect fire effectively in large spaces [31]. Additionally, deep learning algorithms display eminent detection performance in forest fields and can strongly detect objects in environments obscured by street trees, lighting changes, occlusion, shadows, and structures within the forests.

In 2016 EFFIS reported that over 54,000 wildfires were reported across Europe consuming almost 376,000 hectares. The values indicated a decline in forest fire incidents by about 20% compared to the wildfire incidents reported between 2006 and 2015 [9]. The percentage reduction can be attributed to technological advancements in early wildfire detection. Nevertheless, wildfires remain a major problem that claims properties, and early detection is necessary to prevent them. In 2022, the National Interagency Fire Centre in the United States documented 50 significant forest fire incidents between January 1 and June 29. The number surpasses the 10-year average with approximately 192,016 acres blazed [35]. In recent times, wildfire detection has witnessed immense advancements through the use of the latest technology [9]. The latest technology used in wildfire detection and monitoring include unmanned aerial vehicles (UAV), sensor nodes or wireless sensor networks (WSN), spacecraft technique, high-tech sensor and camera devices, and carbon (IV) oxide technique.

The Internet of Things (IoT) serves as a crucial model, addressing challenges in health, transportation, security, robotics, and agriculture effectively and dependably [1, 2]. IOT devices can sense, communicate, and process data therefore offering optimal connectivity that can be utilized in monitoring, controlling, and automation [11]. IoT-based platforms are currently considered for disaster management due to their attractive features such as flexibility, interoperability, heterogeneity, and lightweight [45]. In the wildfire context, it is important to detect the wildfire's exact location. A streamlined IoT platform for wildfire management is anticipated to bring significant economic, social, and environmental impacts to society. Nevertheless, IoT platforms for disaster management require robust, efficient, and dependable communication among IoT devices [15].

While IoT networks are anticipated to support millions of IoT devices, insufficient infrastructure over forests and limited IoT devices' power and complexity make data aggregation unachievable using standard IoT networks. To solve this challenge, unmanned aerial vehicles (UAV), also known as drones or Unmanned Aerial Systems (UAS) may be used [18]. UAVs are systems or vehicles that are operated remotely and travel by flight [9]. The data collected by UAVs are commonly accurate, in real-time, and give unique vantage points that would be dangerous, inaccessible, and time-consuming to acquire by emergency responders. The collected data often take the forms of GPS location, video feeds, images, and sensor node readings. The UAV is controlled remotely by automated systems or humans [44]. UAVs support augmented data rates and dependability demands for cellular communication networks. Additionally, UAVs are flexible and cheap making them suitable for reaching remote and dangerous areas for disaster recovery [6].

Sensor nodes consist of gases, temperature, and humidity to monitor the environment for fire and make alerts [7]. In the US various sensor nodes have been designed to detect wildfire early. In 2006, the FireWxNet sensor node was designed with relative humidity, temperature, wind direction, and wind speed sensor types, whose source of power was four batteries and solar [9]. In 2020, BurntMonitor sensor nodes were designed whose sensor types were temperature and humidity. In 2022, N5 sensors were designed with an IR camera, proprietary nanowire-based gas sensor array, and particulate matter detector, whose source of power was a rechargeable 30,000mAh battery and solar panels [35].

Stationary camera networks comprise advanced and feature-rich, interconnected cameras that keep a check on a huge area for fires. Initially, camera networks comprised camera videos and images streamed to a control room, where a technician would manually scan the feeds for fire signs. Currently, the camera networks are still the main system drivers, but are regularly partnered with other systems such as communication servers and AI, to fully optimize the camera to the preferred area. In the United States, the camera networks have been widely implemented. For instance, ALERTWildfire is a combined effort between the University of California San Diego, The University of Nevada (Reno), and the University of Oregon. In the western region of the United States, hundreds of ALERTWildfire systems have been deployed to detect and monitor fires. In 2021, PG&E (Pacific Gas & Electrical) put in place ALERTWildfire in central and northern California in partnership with Alchera, an AI company. Pano AI, a San Francisco-based company used AI on HD camera video feeds to detect wildfires automatically and minimize the response time to fire.

Artificial intelligence plays a crucial role in wildfire management, from detection to remediation [63]. Merged with remote sensing data, AI can create forest distribution maps, which include oil composition, topography, density, and tree species to predict wildfire risks [47]. The growth of AI has led to the emergence of models like deep learning, machine learning, and CNN. Machine learning-based fire detection algorithms depend on manually extorting visible data from images. These features only focus on the superficial features of the flame, which could lead to information loss when extorting manually [31]. Unlike machine learning algorithms, deep learning automatically extorts and familiarizes complex feature representations [47]. "CNN-based models utilize frames from surveillance systems as input, and the predicted result is sent to an alert system [44].

Forest fire detection and monitoring utilizes a variety of systems, methodologies, and sensors to improve on early detection, response, and management of wildfires. Remote sensing plays an important role in wildfire detection and monitoring [7]. It entails the use of aerial photography, satellite imagery, and other sensor technologies to gather real-time data about fire incidences, smoke plumes, fire hotspots, and burned areas. Remote sensing facilitates the identification and tracking of wildfires and gives helpful information for resource allocation and decision-making. Geographic Information Systems (GIS) merges spatial data for resource allocation and risk mapping. On the other hand, weather prediction models and monitoring systems help in early warning systems and forecasting fire weather [13]. Fire detection systems such as satellite-based and ground-based sensors establish heat signatures, smoke, and flames for quick response. Sensor networks constantly keep an eye on environmental conditions, while artificial intelligence and machine learning evaluate data for fire detection algorithms. Fifth generation (5G) technology is anticipated to increase the speeds of data transfer from 1Gbps to 20Gbps. 5G technology users will access information and data rapidly as due to innovation [57]. This is an important advancement, especially for emergency services, military, and urgent response teams. Nevertheless, improved solutions will be required since the battery life of the 5G-enabled devices will experience significant losses because of the utilization of high-powered signal boosters. Due to barriers such as buildings, 5G users will require more 5G radios in urban settings [22]. On the other hand, the technology is still insufficient within the rural settings. The growth of communication technologies and the spread of intelligent mobile devices has greatly helped in the development of 5G technology. In the construction industry, 5G technology will not only enhance a building's intelligence but will also accelerate its advancement [2]. Therefore, 5G technology will be a technical facilitator in industrial uses and economic opportunities.

## 2. LITERATURE REVIEW

### 2.1. Proactive detection and prevention of wildfires using advanced technology.

In the US, some of the deadliest wildfires such as California's 2018 wildfires resulted from power systems. The growing demand for power in the US requires transmission lines that cover a long distance and have large capacities. Some transmission lines pass through fire-threat areas such as forests, thus escalating wildfire risk. To counteract these wildfires, energy firms develop wildfire mitigation plans. According to [37]. Minimizing wildfire ignition along transmission lines in both rural and urban areas involves employing strategies such as vegetation management. Vegetation management includes activities such as vegetation removal, pruning, application of herbicides, and inspections. However, advances in technology such as aerial imaging e.g., LiDAR and drone (UAV) technology enable inspection and patrol of energy grids. The technologies also enable power suppliers to identify areas that need trimming thus improving situational awareness and conducting more effective condition-based trimming [37]. As a result, vegetation management crews can be dispatched more effectively to address the most vulnerable areas.

Since wildfire ignitions caused by humans are preventable, raising public awareness and education could be vital in minimizing the number of large wildfires as community encroachment increases [20]. Public agencies primarily hold the responsibility for wildfire prevention, with community organizations, non-profits, and emerging partnerships playing supplementary roles [21]. Public service announcements and education campaigns disseminate knowledge to reduce wildfire risks through various channels: social media platforms like Twitter, YouTube, and Facebook; traditional mediasuch as radio and technology, print materials, school campaigns, and websites. Dissemination of information to the public increased awareness of wildfire risks. Educational programs are designed to not only give information on wildfire prevention strategies but also to influence the perception of wildfire risks, attitudes concerning different prevention strategies, and beliefs about effectiveness. In Canada, FireSmart is a program utilized to minimize the susceptibility of private property and communities to fire risk. The program was initially initiated to create awareness and solutions that were workable for vulnerable communities [17].

## 2.2.  Advanced response mechanism to wildfire using drones and IOT devices.

In today's world, the utilization of IoT in military applications has become a necessity due to increased anti-military activities, which have become a threat to many nations [19]. IoT provides solutions to military threats by transferring information in a faster, better, and safer manner with the aid of reliable and powerful wireless communication. [26] evaluated the application of IoT in military operations in a smart city by means of situational awareness in critical situations. An alliance nation may be faced with a disaster situation, which can impact the city's population. Therefore, situational awareness is vital so that resources such as supplies and personnel may be prioritized to help those in most need. Situational awareness can significantly be improved through information acquired from IoT devices. The information acquired is given to the military through warnings and signals. Statistical analysis of information gives a probability of an upcoming problem and its solution.

The utilization of UAVs in the photogrammetry and remote sensing (PaRS) area has become popular [52]. UAVs can acquire information using on-board infrared automatic cameras or visual cameras. However, merging more UAVs results in enhanced fire monitoring services, such as a complementary view of wildfires, big fire coverage, and fire severity assessment [38]. UAV-based remote sensing is utilized in farming and forests enabling decision-making. However, improved data pre-processing with various spatial and temporal data handling software applications is needed. Software can be utilized efficiently to make better decisions in the future [40]. On the other hand, UAVs with deep learning methods can be utilized in fire detection and monitoring as it results in enhanced disaster modeling when merging geo-tagged events that are utilized in geospatial applications [28]. Deep learning is efficient in high-level learning; nevertheless, significant training results in optimal results.

## 2.3.  Active monitoring and mapping with remote sensing and signaling leveraging on 5G technology.

With the increasing damage caused by wildfires, their effective and scientific prevention and control have attracted attention worldwide. The development of remote sensing technologies executed in monitoring early warning and fire spread has become the direction for their prevention and control [58]. Nevertheless, a single remote sensing data gathering point cannot concurrently meet the spatial and temporal resolution requirements of wildfire spread monitoring [33 58]. For example, a study was conducted in Muli County and Sichuan Province, China to monitor wildfire multi-source satellite remote sensing image data from Landsat-8, MODIS, Planet, GF-1, Sentinel-2, and GF-4. The remote sensing data allowed for the rapid and efficient acquisition of the fire spread time series at different intervals. Fire severity and fireline information were obtained based on the computed differenced normalized burn ratio (dNBR). The study gathered the terrain, meteorological, human, and combustible factors related to the fire. The collected data was analyzed using the random forest algorithm [58]. The study results indicated that multi-source satellite remote sensing images could be used and executed for time-evolving wildfires, allowing firefighting agencies and forest managers to organize improved and timely firefighting actions and escalate the efficacy of firefighting strategies. [58] Compared to the single remote sensing image, multi-source remote sensing images are of low cost, efficient, and accurate as they play a vital role in identifying fire spots and extracting burned areas.

Emergency alert systems (EAS), like the one implemented in the United States in 1997, have limitations [29]. For instance, current systems sometimes fail to reach individuals inside buildings due to poor reception or are ineffective for location-specific emergencies. Modern alert systems are now using Information and Communication Technology, including broadcast and cellular systems. In this context, CBS (cell broadcast service) is a significant component used by cellular networks to broadcast emergency alerts to all users in a particular area at once. On the other hand, sirens may provide quick alerts but fail to achieve the desired results as the sound may fail to reach all locations while some people may ignore the sirens. Networks such as WiFi and Ethernet are susceptible to failure during emergencies due to potential power outages leading to network shutdowns causing communication failure.

To avoid such inconveniences during emergencies, modern EAS utilizes ICT such as broadcast and cellular systems [25]. The cellular system provides a cell broadcast technology that delivers emergency alerts to all users in particular cells simultaneously. The cell broadcast mechanism is referred to as CBS (cell broadcast service) in the third-generation partnership project (3GPP) standard group. The 3GPP specifies the CBS protocols for 2G/3G/4G/5G cellular systems [48]. Whenever an alerting authority stipulates an emergency area, an alert message is broadcast by the base stations to all cell users using the CBS protocol. Further, the CBS enables rapid text delivery due to concurrent transmissions. However, traditional CBS has limitations such as delivering only text-based messages, which might not be intuitive or accessible for everyone [11]. Also, in the United States, the EAS based on CBS is referred to as wireless emergency alerts (WEA) [7]. The WEA system also displays latency and text-based message weaknesses.

To address these shortcomings, a new approach in 5G cellular systems has been proposed [11]. This novel method incorporates images into alert messages, making them more universally understandable and quicker to convey, especially beneficial for those who might not be literate or familiar with the local language. This enhancement in 5G uses reserved bits in messages to embed image codes without reducing text content, optimizing alert clarity and efficiency.

## 3. METHODOLOGY

This research relied on secondary data sourced from government databases including weather data, Topographical features, Vegetation data, land use, and historical fire data. Other information like human activities, remote sensing, meteorological forecasts, underground sensors, and geospatial features were also collated where possible. This data is preprocessed and analyzed through exploratory data analysis. Using Python and R programming, a machine learning algorithm is implemented to train and develop the best model for predicting possible wildfire's early stages. The detailed steps include feature engineering, data split into training and test data, model selection, training and evaluation, hyperparameter tuning, and model validation. Some of the Machine Learning Models explored include Support vector Machine, Deep learning Model (CNN and RNN), Random Forest, Gradient Boosting Machine, Gaussian process Regression, and ensemble methods.

In addition, past publications were reviewed through content analysis, and using narrative synthesis to present the observations from various studies, including the relevance of 5G, IoT devices, and drones in wildfire monitoring and management.

## 4. Exploratory Data Analysis

### 4.1. Wildfires in the United States

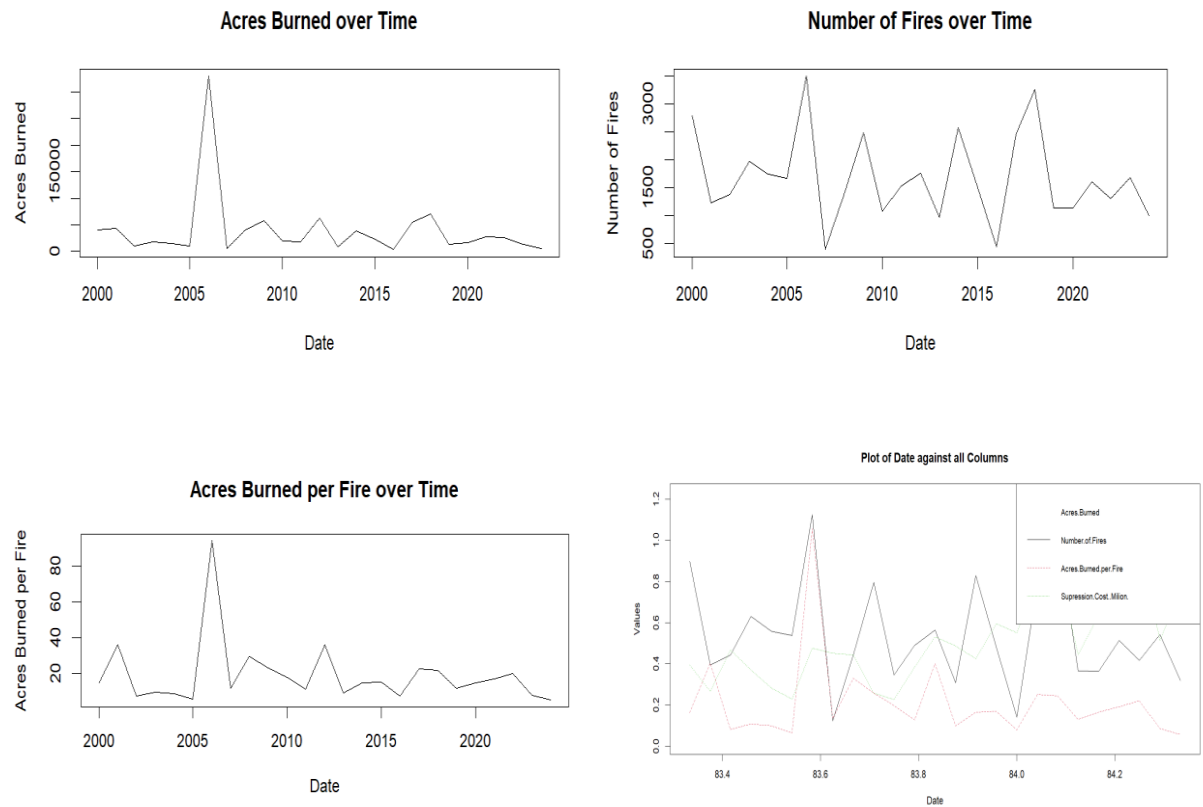

**Fig 1-4: Plots of wildfire frequency and acres burned in the US between 2000 and 2023**

The plots show details of wildfire across United States between 2000 and 2023 including the number of fires, acres Burned over time, and specific acres burned per fire. Its output shows close to 200k fires, with over 18m acres burned. There were occasional spikes in several years and more recently some notable reduction from the year 2000. This may be mainly due to the suppression efforts, indicative of the huge amount of funds invested in wildfire management.

The economic cost of wildfires over the measured period is estimated to be approximately $350 billion. Annually the wildfire costs range from $7.6 billion to $62.8 billion while the yearly losses from $63.5 billion to $285 billion [56]. Data on the suppression cost of wildfires shows that there has been a steady increase in the money spent on suppressing wildfires. Despite the significant cost and reduction in fire incidence lately, there are still risks of fire incidence across several states in the United States.

## 4.2 Proactive Detection and Prevention of Wildfires Using Advanced Technology

Proactive detection and prevention of wildfires is enhanced by advanced technology, which makes it possible for timely intervention. [7] pointed out that incorporating ICT systems into the environment can improve the environment with more features including self-monitoring and self-protection abilities which gives the environment some level of intelligence. This will enable the environment to become an intelligent self-monitoring, self-protecting, and self-aware environment. The environment can promptly respond to changes and alert relevant authorities in real-time, enabling them to take appropriate actions to prevent significant incidents like wildfires. Figure 5 illustrates an overview of fire monitoring and detection methodologies.

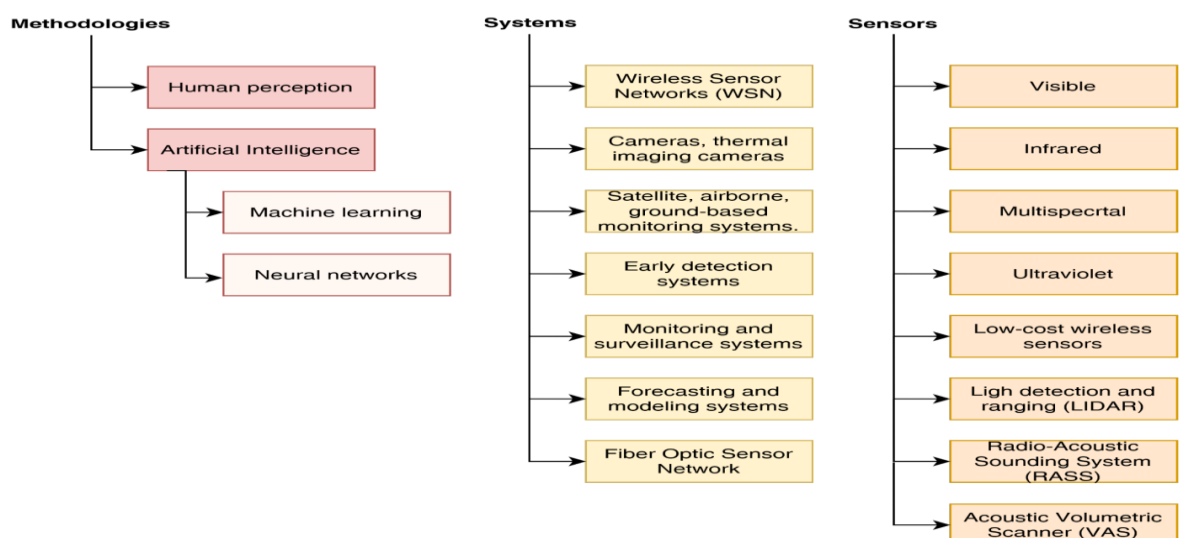

Figure 5: fire monitoring and detection methodologies overview. Adapted from [7]

**4.2.1 Automated Fire Prevention**

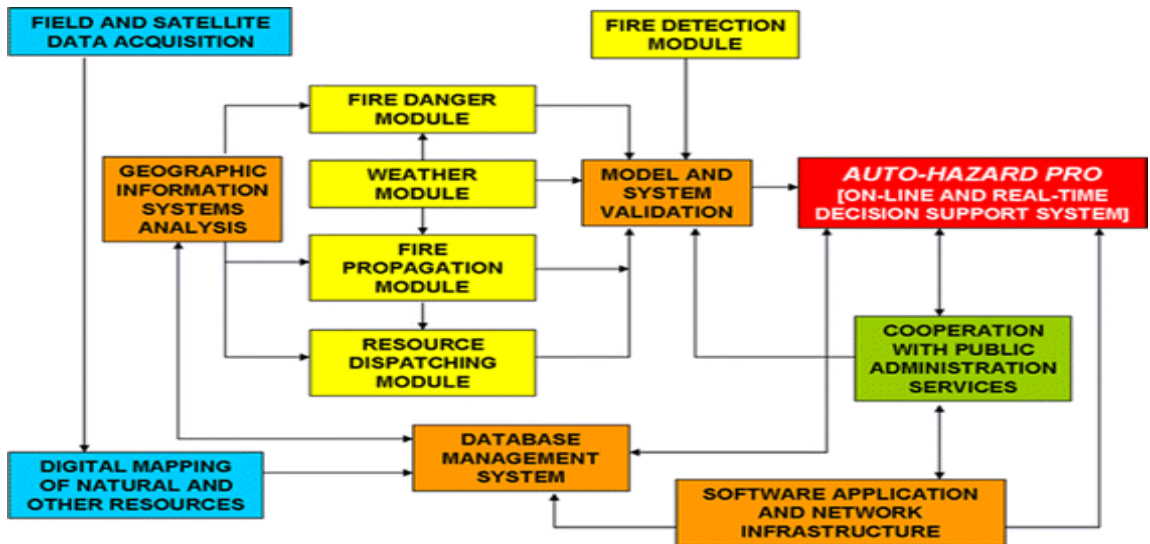

Figure 6: DSS component of the automated fire prevention. Adapted from [22]

A decision support system (DSS) presents an opportunity to help fire management authorities in decisions that prevent fire incidents. The Automated Fire and Flood Hazard Protection System (AUTO-HAZARD PRO) is a Decision Support System (DSS) that encompasses proactive planning, weather data management, a geographical data viewer, predictive risk assessment, fire spread modeling, automated fire detection, strategic resource allocation based on relevant principles, and real-time emergency management during fire incidents. [27]. Figure 6 illustrates the DSS component of the automated fire prevention developed in Europe.

**4.2.2 Wildfire Modeling**

Wildfire modeling simulates wildfire to understand and predict wildfire behavior in an effort to support wildfire suppression, enhance firefighters' and public safety, and lessen damages caused by fire [59]. The Fire Modeling Services Framework (FMSF) is an example of a wildfire modeling system that can predict flame lengths, rates of fire spread, and fire progression. It incorporates the following application (Table 1) that works towards proactive fire prevention and management. Figure 7 is a representative view of the (FMSF) and hosted models/tools.

**Table 1: Fire Modeling Services Framework Applications**

| Application | Function |
|---|---|
| FlamMap | A fire analysis desktop application. |
| MTT (Minimum Travel Time) | The MTT algorithm computes 2-dimensional fire growth. |
| RANDIG (Random Ignition) | A probabilistic 2-dimensional fire spread model. Quantifies the relative likelihood and intensity of fire. |
| FARSITE (Fire Area Simulator) | A fire growth simulation model. Automatically calculates fire growth and behavior. |
| FSPro (Fire Spread Probability) | A strategic decision aid tool examining fire progression threat as informed by uncertainty in the fire environment. |
| Spatial FOFEM – Consumption and Emissions | An application that forecasts immediate fire consequences, such as fuel consumption, soil heating, smoke emissions, and tree mortality. |

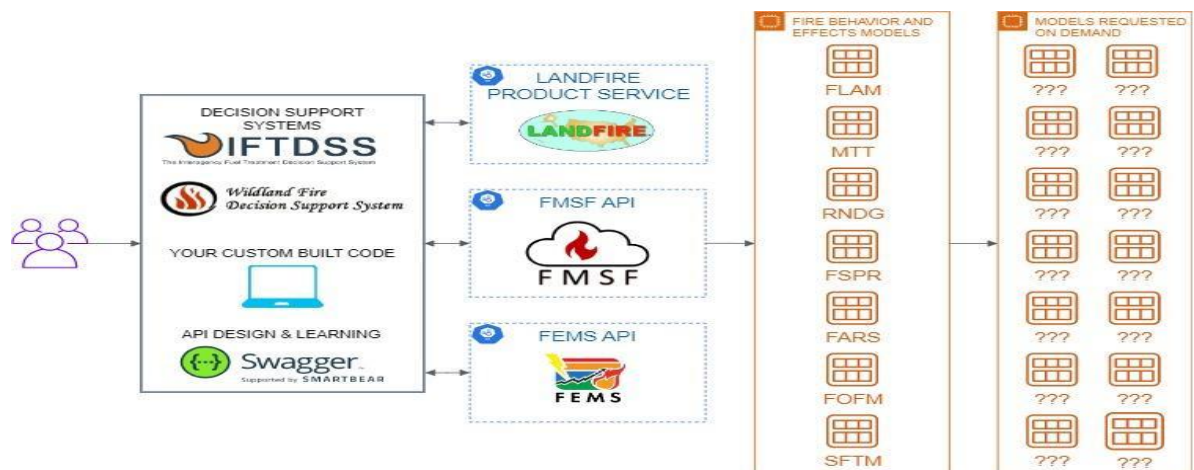

**Figure 7: Fire Modeling Services Framework (FMSF) . Adapted from [49]**

### 4.3 Advanced response mechanism to wildfire using drones and IOT devices.

Drones and IOT devices present great opportunities for responding to wildfires due to their advanced capabilities. [54] presented a model of IoT and drone-based for detecting and responding to forest fires. This model works by deploying IoT-based sensors on trees, grounds, and animals that collect data and transmit it to the control room to put out the fire. The animal sensor is deployed to the animal's body where it detects body temperature and behavior. On the other hand, drones are deployed from the control room whenever the sensors communicate the possibility of a fire. The drones are used for visual confirmation of the fire. In case of fire, the drones will examine the fire intensity and relay back information to the control room for prompt decision to respond to the fire before it spreads. Figure 8 illustrates the system architecture of IoT and drone-based fire monitoring and response systems.

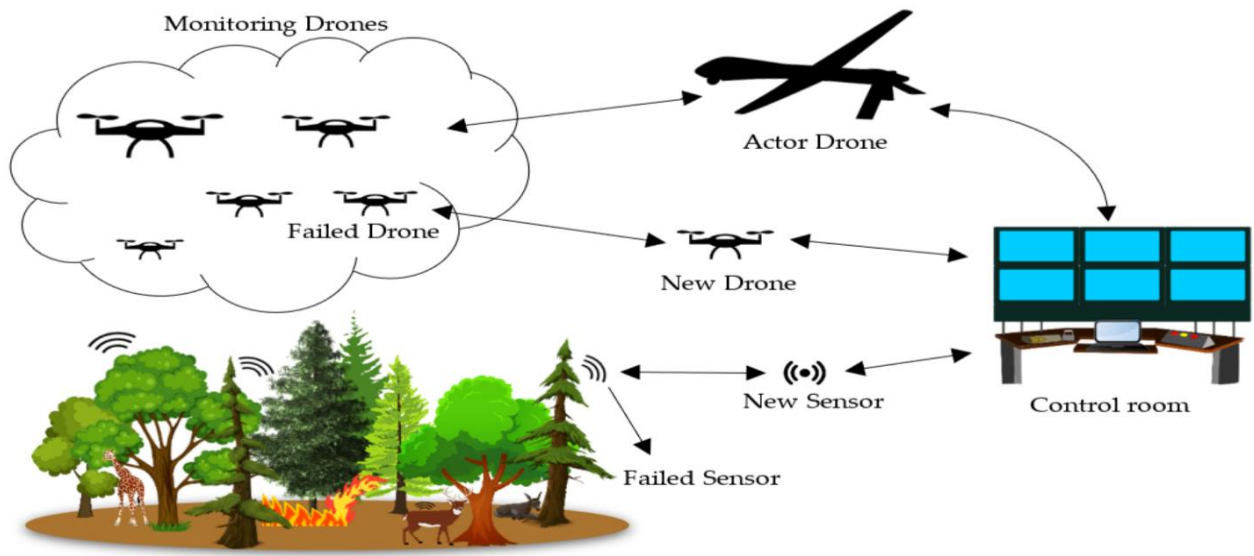

Figure 8: System architecture [48]

## 4.4 Active monitoring and mapping using remote sensing and signaling leveraging on 5G technology.

Fifth-generation technology is meant to handle crucial communication. Ultra-reliable and low latency communication (URLLC) in 5G technology is a feature suitable for vital communication needs, including remote operations with unmanned aerial vehicles (UAVs), robots, and communication among autonomous cars. [60]. [50] presented a UAV-based framework that uses a 5G network to analyze data to detect forest fires. The UAV system is a data collection tool fitted out with different sensors to realize searching and geo-information collection in a single flight. It provides real-time monitoring and the ability to support search and rescue operations in wildfires. The UAVs are deployed in a 5G network to cover the target area and detect fire incidents.

**Table 2: working of the UAV-based framework.**

| Stage | Operation |
|---|---|
| Stage 1 | The map of the area to be scanned is designed, the Data type is decided and the request for the search and rescue operations and region is mapped for UAV operation. |
| Stage 2 | This is the operation stage. The UAV takes off to scan the region and transmit data to the base station in real-time. There are multiple communication links among the UAVs and ground stations as well as the satellite |
| Stage 3 | This is the analysis stage of the real-time data collected. The high-resolution images are transmitted to the base station which monitors for the event detection and coordinates are transmitted in case of an event. |

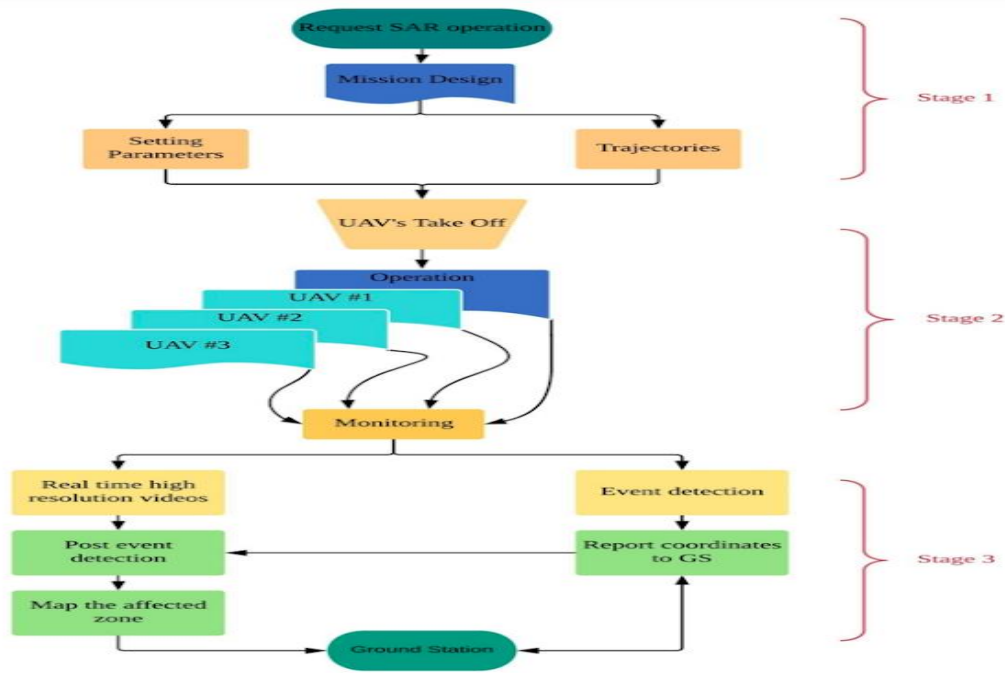

Figure 9: operation of the proposed system for the event detection. Adapted from [44]

## 5.0 DISCUSSION

Wildfire has a significant economic effect, and its management is complicated, dynamic, and rife with incentive problems [4]. According to [4] wildfire can spread very fast threatening people in their path by posing a significant threat to life, property, and local economies. Wildfire also diminishes the quality of air which can negatively affect human health. Advanced technology facilitates proactive detection and prevention of wildfires as it transforms the environment into an intelligent environment. An intelligent environment can self-monitor, self-protect, and be self-aware. This intelligent environment is made possible through Artificial intelligence and Machine Learning models developed for effective wildfire proactive detection and management. With these capabilities, the environment can react to changes promptly and alert the response team in real-time, to prevent major casualties. According to [7] early detection and rapid response systems play an important role in monitoring and surveillance, facilitating timely action. Automated fire prevention can be aided by a decision support system that supports firefighters in making decisions regarding the management of fire.

Technology also offers advanced response mechanisms to wildfires through drones and IoT devices. The IoT-based sensors on trees, grounds, and animals collect data and transmit it to the control room. According to [11] IOT devices can sense, communicate, and process data leveraging 5G technology, therefore offering optimal connectivity that can be utilized in monitoring. IoT-based platforms are currently considered for disaster management due to their attractive features such as flexibility, interoperability, heterogeneity, and lightweight [45]. In addition, drones are used for visual confirmation of the fire whenever an IoT-based sensor reports a potential fire. In the event of a fire, drones examine the fire intensity and relay back information to the control room for prompt decision to respond to the fire. According to [38] UAVs which are also known as drones can acquire information using on-board infrared automatic cameras or visual cameras. Merging more UAVs results in enhanced fire monitoring services, such as a complementary view of wildfires, big fire coverage, and fire severity assessment. Furthermore, UAVs with deep learning methods can be utilized in fire detection and monitoring as it results in enhanced disaster modeling when merging geo-tagged events that are utilized in geospatial applications [28].

Fifth-generation technology supports communication remotely. In wildfire management, 5G technology can enable active monitoring and mapping using remote sensing. The UAVs in wildfire monitoring are deployed in a 5G network to cover the target area and detect fire incidents. According to [9], 5G technology increases the speeds of data transfer from 1Gbps to 20Gbps. The use of 5G technology enables users to access information and data rapidly due to innovation. This is an important advancement, especially for emergency services and urgent response teams including firefighters.

## 6.0 CONCLUSION

The analysis of two decades of wildfire data in the United States highlights the ongoing risk despite recent reductions in incidents. This research underscores the crucial role of cutting-edge technologies such as AI, ML, and 5G in proactive wildfire management. Automated fire prevention systems, guided by decision support technology, improve decision-making processes for fire management, effectively preventing incidents. Wildfire modeling tools aid in understanding and predicting fire behavior, supporting suppression efforts, and ensuring public safety. IoT devices and drones provide real-time data collection, enabling swift detection and response, while 5G technology revolutionizes communication, ensuring rapid information access for emergency services. This technological integration represents a transformative leap in disaster response capabilities, empowering authorities to proactively combat wildfires, ultimately safeguarding lives, properties, and the environment from these devastating natural disasters.

## 7.0 RECOMMENDATION

This research suggests that the government should incorporate advanced technology to prevent and control wildfires. Implementing AI and ML can create a self-monitoring and self-protective intelligent environment, effectively controlling the occurrence and propagation of wildfires. Additionally, integrating 5G technology for communication networks in wildfire-prone areas can facilitate the collection and transmission of data related to wildfire threats. To ensure the safety of responders during active wildfires, unmanned aerial vehicles (UAVs) equipped with 5G technology and AI capabilities should be deployed for rescue and firefighting operations. Furthermore, the study advocates for the development of Decision Support Systems (DSS) to aid firefighters in detecting potential wildfire outbreaks and making informed decisions regarding wildfire management strategies.

## AUTHORS

***Stanley Okoro*** *is a highly skilled and innovative ICT Engineer with expertise in 5G, Cloud Computing, Data Analytics, Generative AI, and ML. He has a bachelor's degree in Electrical and Electronics Engineering, an MBA in Global Business Management, and is currently completing his second master's degree in Computer Science and Quantitative Methods at Austin Peay State University, Clarksville Tennessee, USA. He has a deep passion for cutting-edge technology and specializes in bridging business objectives with technical solutions.*

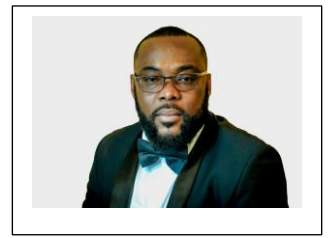

***Alexander Toluwani Lopez*** *is a seasoned data professional with a diverse academic background and over 5 years of experience. He holds a Bachelor of Science in Electrical & Electronics Engineering from the University of Lagos, Nigeria, and is pursuing a master's in computer science and quantitative Methods with a focus on Predictive Analytics at Austin Peay State University, The USA. His expertise in data analysis and software development, coupled with a strong passion for data analytics and machine learning makes him an asset in the field of data science.*

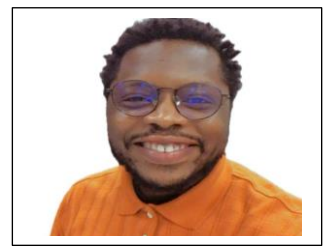

***Austine O. Unuriode*** *has a bachelor's degree in mathematics (Nigeria). He is currently pursuing a master's degree in computer science and quantitative methods, with a concentration in database management and analysis (USA). Austine has over 5 years of working experience as a data and business analyst. He has a keen interest in data analytics and data engineering, particularly in cloud computing.*

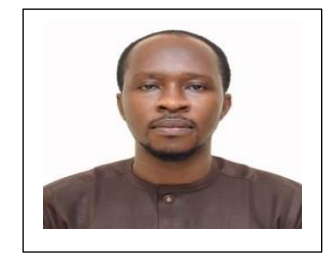